\documentclass[%
 reprint,
nofootinbib, 
 amsmath,amssymb,
 aps,
]{revtex4-2}

\usepackage{graphicx}
\usepackage{dcolumn}
\usepackage{bm}
\usepackage{braket}
\usepackage{simpler-wick}

\begin{document}

\preprint{APS/123-QED}

\title{Redundancies\\and\\Profundities}

\author{Kyle Singh}\thanks{kyle.singh.gr@dartmouth.edu}
 \affiliation{
 Guarini School of Graduate and Advanced Studies\\
64 College Street
Anonymous Hall Suite 102
Hanover
New Hampshire 03755-3563
}%

\date{\today}

\begin{abstract}
We reevaluate the status of the gauge principle and reposition it as an intermediary structure dependent on the initial conditions we endow on our theory. We explore how the gauge symmetry manifests in the context of basic quantum electrodynamics, spontaneous symmetry breaking and the modern scattering amplitudes program. We also investigate the addition of an auxiliary field in $\phi^4$ theory and see how the dynamics are altered. Modal language is pointed to and utilized as a convenient way to articulate the weight gauge symmetry demands in our theories as well as the principles of locality and Lorentz invariance. A shifting scale ontology is introduced with regards to the gauge  principle and other structures of Quantum Field Theory in general. 

\end{abstract}

\maketitle


\section{\label{sec:level1}Introduction }
The gauge principle is widely regarded as the cornerstone of fundamental physics. However, the modern scattering amplitudes program, as well as the discovery of the AdS/CFT correspondence, have highlighted the redundancy that comes from the implementation of the symmetry in Quantum Field Theory and in general. Moreover, independent of such recent developments the ontological status of the gauge principle has been brought into question. This places the gauge principle in a unique position. It is a concept that is fundamental; one that no doubt leads us to profound physical insight by fixing the structure of fundamental interactions in the standard model while also revealing previously hidden degrees of freedom, elementary particles, that are present in nature. However, in other regimes, its presence masks the simplicity underlying our theory and certain physical aspects of our systems. We need to find a way to categorize that which is redundant and also profound in our physical theory. 

More generally, our basic theories are such that intermediary concepts, as we will call them,  map onto physical data and are often most important. Consider, for example, the fact that in General Relativity one is free to make a choice of coordinates based on what the physical system demands. Particular coordinate choices, such as in the distinction between the Schwarzschild and Kruskal coordinates for black holes, reveal different aspects of the spacetime structure. In this case, for example, the Kruskal coordinates resolve the non physical coordinate singularities endemic to the Schwarzschild geometry. Such freedom can easily allow us to question the role of a background space itself. Here space is instead in an intermediary position. It is fundamental as a background but wholly arbitrary based on the physical system and initial conditions we impose. 

Not only is gauge symmetry profound in our fundamental theories, it is unavoidable. Carlo Rovelli establishes the ubiquity of gauge and claims that "Gauge invariance is not just mathematical redundancy; it is an indication of the relational character of fundamental observables in physics" (Rovelli, 7). He sets up a coupled dynamical system and shows that gauge variables are necessary in order for one to capture the full dynamics. In other words, one can not decouple the system without the presence of gauge variables. \footnote{See Ref. [2] for a full discussion of this.} For Rovelli, the gauge symmetry reveals the "relational structure of our world", as gauge interactions describe relative quantities with more than one object (Rovelli, 7). For example, consider the fact that in ordinary quantum mechanics, we can only measure differences in energy.  Rovelli insists that treating gauge as pure redundancy ignores this relational structure. 

Indeed, claiming that the gauge principle results in pure redundancy or that it is somehow inherent to our fundamental theory does not capture the unique position it operates within. We are then left with a view of the gauge principle that is purely intermediary. Our options for how we should treat the gauge symmetry were aptly categorized by Michael Redhead in what has now come to be known as Redheads trilemma. Redhead purports that we have three options with how we choose to treat the gauge principle. We can either claim that gauge symmetry is physical and motivate physical structures directly representative of gauge fields, try to reformulate the entire theory in terms of quantities that are purely gauge-invariant, or we can let non-gauge-invariant quantities  enter as surplus structure and develop the theory accordingly, adding further surplus structure if necessary to make the theory work. 

The third option is one often taken by physicists for practical purposes and is the stance we will undertake. The question then becomes one in which we ask ourselves how we should categorize such intermediary concepts in theoretical physics and more broadly in mathematical representation. Redhead's three propositions do not fully articulate such a position, rather it seems that they capture some aspects of how we wish  to treat the gauge symmetry in each of them. His distinctions seem to arise within the context of a particular theory or set of calculations. For example, with respect to his second proposition, we can look to Wilson loops which are completely gauge invariant quantities and work out the dynamics of our QFT in terms of them, however they lead to non-local physics, as is well evidenced by the Aharonov–Bohm effect. 

Indeed, parsing out various physical systems and regimes of our QFT and observing what role the gauge principle has within them will be central to how we choose to speak about its status. Furthermore, this will allow us to set up a shifting scale ontology and classify various pillars of our theory in an ontological way. 

Not only do we need to deal with this status of gauge symmetry, we must also find a way to incorporate its inherent ambiguity both in its implementation and with respect to the physical objects the symmetry comes to represent. We do not directly discuss surplus structure broadly in mathematics, as Guay has suggested we ought to do; however it is not so much of a concern in the particular way we choose to position gauge symmetry since we are teasing out the physical data that it leads us to and treating its inherent redundancy as a triviality since we can easily choose a particular gauge even though we are given infinite choices from which we can begin our computations. The redundancies will only be important if they mask the underlying simplicity of our theory. This will be discussed within the context of the modern scattering amplitudes program and  it is within this context that the discussion of surplus structure in general may be apt. 

We now begin with a brief summary of the gauge principle and its operation in the context of electrodynamics. 

\section{\label{sec:level1}The Gauge Principle }

Let us briefly review the gauge principle in the context of classical electromagnetism. We begin with the Lagrangian for complex scalar field theory 
\begin{equation}
    \mathcal{L}=\left ( \partial_\mu\psi \right )^\dagger\left ( \partial_\mu\psi \right )-m^2\psi^\dagger\psi
\end{equation}
The Lagrangian possesses a U(1) symmetry by the following replacement
\begin{equation}
\psi(x)\rightarrow\psi(x)e^{i\alpha}
\end{equation}
Note that this symmetry is a global one, namely the field is changed by the same amount at every spacetime point. As is standard, we can then work out the consequences if we impose the notion that our Lagrangian remain invariant under local transformations with the following replacement
\begin{equation}
\psi(x)\rightarrow\psi(x)e^{i\alpha(x)}
\end{equation}
Clearly, under such a prescription the Lagrangian is not invariant. In order to restore the local symmetry, we introduce a new field $A_\mu(x)$ which cancel out the extra terms resulting from the requirement of local gauge invariance. We must require that $A_\mu$ transforms as $A_\mu(x)\rightarrow A_\mu(x)-\frac{1}{q}\partial_\mu\alpha(x)$.\footnote{Here, $q$ is the coupling strength and in this case is just an extra parameter which will have physical significance in other theories.}We then introduce the covariant derivative defined as follows
\begin{equation}
D_\mu=\partial_\mu+iqA_\mu(x)
\end{equation}
Given this set of transformations, our theory is locally gauge invariant. The introduction of an additional field in order to obtain local gauge invariance is a ubiquitous feature of gauge theories. In the context of electromagnetism, we utilize the following Lagrangian. \footnote{Note, that this is the same as writing $\mathcal{L}=-\frac{1}{4}F_{\mu\nu}F^{\mu\nu}-J^\mu A_\mu$}
\begin{equation}
\mathcal{L}=-\frac{1}{4}(\partial_\mu A_\nu-\partial_\nu A_\mu)(\partial^\mu A^\nu-\partial^\nu A^\mu)-J^\mu A_\mu
\end{equation}
The equations of motion are the first two Maxwell equations.
\begin{equation}
\partial^2A^\nu-\partial^\nu(\partial_\mu A^\mu)=J^\nu
\end{equation}
We can rewrite the transformation of our gauge field in the following more generalized form
\begin{equation}
A_\mu(x)\rightarrow A_\mu(x)-\partial_\mu \chi(x)
\end{equation}
Of course, in any physical theory we want the fields we introduce to hold physical significance. The function $\chi(x)$ has, for all intents and purposes, an infinite number of symmetries, one for each function.\footnote{This conundrum in itself is indicative of gauge as redundancy. It would be startling to say that presence of infinite symmetries would yield an infinite number of conservation laws following from Noether's theorem, for example! Here two states related by a gauge transformation are indeed the same physical state.  } Therefore, we must take further steps in constraining our gauge field to facilitate its representation of a physical object. In the case of electrodynamics, the gauge field contains the dynamics of the photon. The following two conditions, known as choosing a gauge, are imposed to ensure that the gauge field has two degrees of freedom since the photon can only have two polarizations
\begin{equation}
\partial_\mu A^\mu(x)=0
\end{equation}
\begin{equation}
\partial_0\xi=A^{'}_0
\end{equation}
These particular gauges are known as Lorentz gauge and Coulomb gauge, respectively. The gauge principle then refers to the procedure of introducing this additional dynamical field to maintain local gauge symmetry. Moreover, as we have just seen, the gauge field dictates the form of the coupling.

\emph{Prima facie}, there seems to be no reason at all to impose local gauge invariance. Moreover, it seems quite contrived to insist that the gauge field introduced to preserve our desired symmetry must have a precise set of dynamics correspondent to a particular object in a theory; in other words our designation of the gauge field as a photon was put in by hand and not derived from the gauge principle itself. Even more striking perhaps, is the fact that we can never measure $A_{\mu}$. That nature seems to require us to introduce such objects is truly incredible.\footnote{All of this may be a false problem, of course, and Rovelli aptly calls into question whether or not we should ask our mathematical procedures to have a purpose; asking us whether or not we should conclude that it was the purpose of humans to kill large mammals if we were indeed responsible for their deaths.}

Indeed, Martin calls these difficulties with the gauge principle out and writes that the idea that the gauge principle "'dictates' [or] 'determines' the form of fundamental interactions as well as the existence of certain physical fields must be taken with a large grain of salt" (Martin, 233).  He argues that, at best, the gauge principle should be taken as heuristic and offers a differing approach to the logic of nature, viewing the gauge symmetry as not a fundamental physical principle, rather, as a relic of a theory that is more fundamental, in particular as renormalizable theories, that works in conjunction with other physical requirements such as Lorentz invariance. 

It is clear that a proper revaluation of the status of the gauge principle  in our physical theories is necessary. We will expand on Martin's thesis and seek to make more general modal statements relating the conception of local gauge symmetry to our physical theory as one apparatus. In doing so, we do not contradict Rovelli's argument on the ubiquity of gauge or the arbitrariness inherent to our imposition of the gauge principle in itself. Instead, we reconsider its position as a principle in our physical theory all together.

\section{\label{sec:level1}Pure Redundancy}
Consider the following Lagrangian in $\phi^4$ theory. 
\begin{equation}
\mathcal{L}=\frac{1}{2}(\partial_\mu\phi)^2-\frac{m^2}{2}\phi^2-\frac{g}{8}\phi^4
\end{equation}
We shift the Lagrangian by adding a new $\sigma$ field in the following way\footnote{Shifting our Langrangian in such a fashion is commonly referred to as a \emph{Hubbard-Stratonovich} transformation.}
\begin{equation}
\mathcal{L}'=\mathcal{L}+\frac{1}{2g}\left ( \sigma-\frac{g}{2}\phi^2 \right )^2
\end{equation}
The Lagrangian is then 
\begin{equation}
\mathcal{L}'=\frac{1}{2}(\partial_\mu\phi)^2-\frac{m^2}{2}\phi^2-\frac{g}{8}\phi^4+\frac{1}{2g}\left ( \sigma-\frac{g}{2}\phi^2 \right )^2
\end{equation}
In QFT, the Green's functions of our theory tell us about the dynamics. We can compute these via the generating functional
\begin{equation}
\mathcal{Z}[J]=\frac{\displaystyle\int \mathcal{D}\phi\mathcal{D}\sigma \text{exp}\left [ i\displaystyle\int d^4 x\left ( \mathcal{L}'+J\phi \right ) \right ]}{\displaystyle\int \mathcal{D}\phi\mathcal{D}\sigma \text{exp}\left [ i\displaystyle\int d^4 x\mathcal{L}'  \right ]}
\end{equation}
If one carries out the above functional integral, it is straightforward to show that the expression written above for our newly defined theory is the same as the original one. To emphasize this, we can compute the equation of motion for the $\sigma$ field
\begin{equation}
\frac{\partial\mathcal{L}'}{\partial\sigma}-\partial_\mu\frac{\partial\mathcal{L}'}{\partial(\partial_\mu\sigma)}=0
\end{equation}
and find that 
\begin{equation}
\sigma=\frac{g}{2}\phi^2
\end{equation}
There are no time derivatives present in the computed equation of motion. Therefore, our newly added field does not contribute to the dynamics of this system. Furthermore, we can eliminate the additional field since it can only provide a constraint on the theory. 

We can promote the fields of our system to operators and write down the vacuum-to-vacuum expansion of the $\mathcal{S}$-matrix for our theory. This yields the relevant Feynman diagrams corresponding to various wick contractions of the fields. This further tells us about the role of the $\sigma$ field. The $\mathcal{S}$-matrix operator reads as follows
\begin{equation}
\bra{0}\mathcal{S}\ket{0}=\bra{0}T\left [ \text{exp}\left ( -i\displaystyle\int d^4x \frac{1}{2}\sigma\phi^2 \right ) \right ]\ket{0}
\end{equation}
Writing out the first couple of terms in the perturbation expansion yields
\begin{equation}
\begin{split}
-\frac{i}{2}\displaystyle\int d^4x\bra{0}T\left [ \sigma_x\phi_x\phi_y \right ]\ket{0}+\frac{1}{2!}\left ( -\frac{i}{2} \right )^2 \\ \displaystyle\int d^4xd^4y\bra{0}T\left [ \sigma_x\phi_x\phi_x\sigma_y\phi_y\phi_y \right ]\ket{0}+...
\end{split}
\end{equation}
Contractions of the $\phi$ field yield the standard free field propagator. As we have shown, since the $\sigma$ field does not contribute to the dynamics of the theory, nothing can propagate through it. Contractions such as $\wick{\c\sigma_x \c\sigma_y}$ take two spacetime points and identify them with one another playing the role of interaction vertices in the various diagrams which arise. This calculation is an example of a procedure resulting in a trivial redundancy. It is clearly distinct from the procedure undertaken in incorporating gauge symmetry into our field equations. 

Treating the gauge principle as simply an artifact of pure redundancy does not capture its importance to the dynamics of say Quantum Electrodynamics and to the construction of the standard model. The $\sigma$ field's importance in the constructed example and the role of the gauge field are wholly different in terms of what they represent and what role they play in the theory, although both were introduced in an arbitrary fashion. We must then seek to reposition the status of gauge as it relates to QFT and see what we can say about scientific theories in full generality.
\section{\label{sec:level1}Pure Gauge}
As we have reviewed, any Lagrangian with local symmetry must harbor gauge fields. Consider the following Lagrangian for gauged complex scalar field theory
\begin{equation}
\begin{split}
\mathcal{L}=\left ( \partial^\mu\psi^\dagger-iqA^\mu\psi^\dagger \right )\left ( \partial^\mu\psi+iqA^\mu\psi \right )+\\ \mu^2 \psi^\dagger\psi-\lambda(\psi^\dagger\psi)^2-\frac{1}{4}F_{\mu\nu}F^{\mu\nu}
\end{split}
\end{equation}
It  is crucial to note that the sign of the mass term has been flipped. This allows us to invoke the standard symmetry breaking procedure. We insist, again, local gauge invariance and work in polar coordinates by letting $\psi(x)=\sigma(x)e^{i\theta(x)}$ for a unique ground state phase set at $\theta(x)=\theta_0$. The fact that we can not now change the phase of the ground state both locally and globally means that the symmetry is broken in both regimes. Now we observe, as is commonly done, what physical consequences can be derived by computing the particle spectrum of this system. In polar coordinates 
\begin{equation}
\partial_\mu\psi+iqA_\mu\psi=(\partial_\mu\sigma)e^{i\theta}+i(\partial_\mu\theta+qA_\mu)\sigma e^{i\theta}
\end{equation} 
Where the gauge field is represented in our theory in the following way
\begin{equation}
A_\mu+\frac{1}{q}\partial_\mu\theta\equiv B_\mu
\end{equation}
Therefore
\begin{equation}
(\partial^\mu \psi^\dagger-iqA^\mu\psi^\dagger)(\partial_\mu\psi+iqA_\mu\psi)= (\partial_\mu\sigma^2)+\sigma^2q^2B_\mu B^\mu
\end{equation} 
The Lagrangian becomes
\begin{equation}
\mathcal{L}=(\partial_\mu\sigma)^2+\sigma^2q^2B^2+\mu^2\sigma^2-\lambda\sigma^4-\frac{1}{4}F_{\mu\nu}F^{\mu\nu}
\end{equation}
We now invoke the standard symmetry breaking procedure. The minima of the potential are at $\sigma=\sqrt{\frac{\mu^2}{2\lambda}}$. We break the symmetry by setting $\sigma_0=\sqrt{\frac{\mu^2}{2\lambda}}$and $\theta_0=0$. Expanding the Lagrangian in terms of a new field $\delta$ defined as $\frac{\delta}{\sqrt{2}}=\sigma-\sigma_0$ and ignoring constants yields the following
\begin{equation}
\begin{split}
\mathcal{L}=\frac{1}{2}(\partial_\mu\delta)^2-\mu^2\delta^2-\sqrt{\lambda}\mu\delta^3-\frac{\lambda}{4}\delta^4-\frac{1}{4}F^{\mu\nu}F_{\mu\nu}\\ +\frac{A^2}{2}B^2+q^2\left ( \frac{\mu^2}{\lambda} \right )^{\frac{1}{2}}\delta B^2+\frac{1}{2}q^2\delta^2B^2+...
\end{split}
\end{equation}
Here, $A=q\sqrt{\frac{\mu^2}{\lambda}}$. Breaking the symmetry surprisingly results in our theory containing the massive vector field $B_\mu$. Meanwhile, the massless excitations of the $\theta$ field have disappeared. Note that these excitations were only present in the case of global symmetry breaking. Our theory, which once described two massive scalars and two massless photons now describes one massive scalar and three massive vector particles. Imposing the gauge transformation $A_\mu+\frac{1}{q}\partial_\mu\theta=B_\mu$ removes the Goldstone modes. This removal of the Goldstone mode via the prescribed gauge transformations means that it is pure gauge. 

Local symmetry breaking yields the massive physical degree of freedom while removing the nonphysical massless one. This procedure, as is well known, is crucial to the Higgs mechanism and places the Higgs boson correctly within the standard model. Local symmetry yields new physical insight fixing the redundancy of global symmetry breaking.  This manifestation of gauge invariance is distinct from the pure redundancy discussed with regards to the addition of an auxiliary scalar field, however it is an indication of the intermediary role that gauge symmetry plays in a  particular regime that we wish to probe in the context of our field theories. 

\section{\label{sec:level1}Modal Considerations}
Given the preceding discussions, it is now natural to ask ourselves what we can say about modality with respect to the gauge principle. Modal language, even if used loosely for our purposes currently, gives us a convenient way to categorize the gauge principle in relation to the other mechanisms in our QFTs. In order to make modal statements on gauge symmetry, it seems apt to first take one of Redhead's positions on how we should treat it functionally and proceed from there in addressing its status. That the gauge symmetry results in surplus structure, redundant degrees of freedom, means that only a subset of these degrees of freedom can be recognized as physical representations. We can take Redhead's third proposition that theories should keep the gauge invariance for as long as possible. This means that we do not dispose of the local gauge symmetry and take into account its physical predicative power. \footnote{Again, as discussed earlier, we are not so concerned with this subtly and work with the gauge principle, in many ways, after the question of what gauge to work in has been decided upon.}

This conveniently allows us to split up the gauge symmetry into a local piece and a global one.  \footnote{It is safe to say that use of the word "principle" in gauge principle is misleading and not indicative of how we treat gauge symmetry at large. Although the gauge principle in itself refers to the local gauge symmetry, the use of the word would be better served when looking at the role of gauge symmetry in particular as a whole.  } This factorization allows us to evaluate any claims of necessity independent of one another. 

 Global gauge invariance has physical necessity because it carries through as a symmetry in all of our foundational theories although it is, in most cases trivially a part of our theories. Any principles that are essential to the structure of the theory are metaphysically necessary, for example Lorentz invariance. Local gauge symmetry then takes on an intermediary role. Since it is not wholly essential in our theories which are more fundamental, this will be discussed in the context of the modern scattering amplitudes program, we can posit that it is metaphysically possible. It is possible that local gauge invariance is required for the prediction of a photon coupling to a electron, however it may not be. Perhaps it may carry nomic necessity, however such a claim would require that we know the true origin of why we must impose local gauge symmetry at all. This would mean that it would be attached to some underlying law of nature that we clearly do not know of now. For now, in the context of how we are exploring and seeking to categorize the gauge symmetry, it is enough to make the modal distinction we have made above in the hopes of clarifying how we wish to treat gauge within the broader construction of QFT. \footnote{It would be interesting to explore whether such physical concepts, including gauge, tied to physical objects can be categorized within a more rigorous formal modal structure.}

 \section{\label{sec:level1}Scattering Amplitudes and Simplicity}

The modern scattering amplitudes movement has given us a method to compute  amplitudes in a way that forgoes gauge redundancy all together revealing aspects of a more foundational QFT. The standard polarization vectors responsible for describing redundant massless particle states are replaced by spinor-helicity variables which are trivially gauge invariant. \footnote{For a review of the spinor-helicity formalism refer to [4]}This modern incarnation of the $\mathcal{S}$-matrix bootstrap imposes the fundamental principles of locality and unitarity to determine amplitudes. What one finds is that calculations which were once extremely complicated in the traditional Feynman diagrammatic approach become tremendously simplified and almost trivial, thanks to a cancellation of redundancies. Taking the fermion states to be massless for these calculations, we are working in the high-energy scattering limit which constitutes a theory at a more fundamental energy scale. 

Let us compute the color-ordered 4-gluon amplitude, $A_4[1^-2^-3^+4^+]$, at tree level. Recall that such partial amplitudes are trivially gauge invariant. Utilizing the standard Feynman rule for the 4 gluon vertex allows us to write\footnote{Note that one needs to specify a particular set of reference spinors. We have chosen $q$'s such that the $t$-channel diagram vanishes and all $\epsilon_i\cdot\epsilon_j$'s vanish except $\epsilon_2\cdot\epsilon_3$}
\begin{equation}
A_4=\frac{(-i\sqrt{2}g^2)((\epsilon_1\cdot p_2)\epsilon_2-(p_1\cdot \epsilon_2)\epsilon_1)((\epsilon_3\cdot p_4)\epsilon_4-(p_3\cdot\epsilon_4)\epsilon_3)}{(p_1+p_2)^2}
\end{equation}
Translating this into the spinor helicity formalism, one finds the following expression

\begin{equation}
\begin{split}
A_4=\frac{-2g^2}{\braket{12}[12]}\frac{\braket{12}[34]}{\braket{13}[24]}\\ \frac{\braket{12}[24]}{\sqrt{2}[14]}\frac{\braket{13}[34]}{\sqrt{2}\braket{14}} 
\end{split}
\end{equation}

Applying momentum conservation and simplifying this expression gives us the following simple amplitude 

\begin{equation}
A_4=\frac{\braket{12}^4}{\braket{12}\braket{23}\braket{34}\braket{41}}
\end{equation}

Indeed, one finds the following simple expression for all tree level Yang-Mills amplitudes. \footnote{This can be derived using the standard BCFW recursion relations}
\begin{equation}
A_n[1^+...i^-...j^-...n^+]=\frac{\braket{ij}^4}{\braket{12}\braket{23}...\braket{n1}}
\end{equation}
It is a remarkably simple expression that is fully generalized. In the traditional perturbative formalism computing a seven gluon amplitude would require the calculation of 154 separate diagrams, with the amplitude boiling down still to the result above. Without the extra gauge redundancies clouding the fundamental structure of the scattering amplitudes, we can ask ourselves what principles we are left with. Consider the following ansatz for for 3 particle amplitudes
\begin{equation}
A_3(1^{h_1}2^{h_2}3^{h_3})=c\braket{12}^{x_{12}}\braket{13}^{x_{13}}\braket{23}^{x_{23}}
\end{equation}
Under little group scaling on-shell amplitudes transform in the following way, with helicity $h_i$
\begin{equation}
A_n({\ket{1},|1],h_1},...,{t_i\ket{i},t_i^{-1}|i],h_i},...)=t_i^{-2h_i}A_n(...{\ket{i},|i],h_i}...)
\end{equation}
This fixes the following
\begin{subequations}
\begin{equation}
-2h_1=x_{12}+x_{13}
\end{equation}
\begin{equation}
-2h_2=x_{12}+x_{23}
\end{equation}
\begin{equation}
-2h_3=x_{12}+x_{33}
\end{equation}
\end{subequations}
Solving the system of equations we can rewrite the ansatz as follows
\begin{equation}
A_3(1^{h_1}2^{h_2}3^{h_3})=c\braket{12}^{h_3-h_1-h_2}\braket{13}^{h_2-h_1-h_3}\braket{23}^{h_1-h_2-h_3}
\end{equation}
Now, we can consider a 3-gluon amplitude with the following helicity configuration
\begin{equation}
A_3(g_1^-g_2^-g_3^+)=g\frac{\braket{12}^3}{\braket{12}\braket{23}}
\end{equation}
Little group scaling fixes the form of the amplitude. Moreover, the amplitude is fixed by locality, namely that it is compatible with a term of the form $AA\partial A$ in the Lagrangian $\text{Tr} F_{\mu\nu}F^{\mu\nu} $ and not a term that goes like $g'AA\frac{\partial}{\square}A$.

We are now in position to ask ourselves what we are left with in this high energy theory. We are left with the principles of locality, unitarity, and Lorentz invariance, as outlined in the simple calculations above. Gauge symmetry plays  a trivial role in this regime, where calculations are simplified and where a more foundational structure of our amplitudes, and perhaps our QFT, is revealed. Coupling this insight with our previous modal claims, we can revise and categorize a new ontological status for the gauge principle. 

\section{\label{sec:level1}The Ontological Status of Gauge Symmetry}
We begin with a set of principles and regard them as the basis for our ontology. Then the gauge principle, as is customarily defined, cannot fit into our ontological construction. Instead,  we can take its factorized local piece to be one step removed from the fundamental principles of locality, unitarity and Lorentz invariance. Local gauge symmetry, stated as a principle in the way we utilize it, is a part of our ontology in the theory that is less fundamental, at lower energy scales. It is a projection onto the more fundamental theory that becomes necessary at lower energy and resolves itself by exiting the picture at higher energies.

We have, then, a direct manifestation of Occam's razor at higher energy, whereby the theory seems to become simpler and where our ontological stakes become more defined. Indeed, as we have seen we are left with a set of principles embodied within the higher energy theory that dictate all of its tenets. It is quite likely that such a theory is wholly inaccessible to experiment, as has been predicated by String Theory and its exploration in the past several decades. And so, it is not simply enough to say that our ontology, as determined by principles, should be determined by the theory which inhabits higher energy scales. Instead, there is a sense in which our ontology resolves itself at various scales. The base principles carry throughout the scales. Physics is local for quarks, for baseballs and for nuclei. It must be local even for strings if they exist empirically. Certain principles then, get added to our ontology as we lower our energy scale.  

The gauge principle then, is a principle in the sense that its imposition seems to be necessary to obtain the relevant physical phenomena in the effective renormalizable theory and thus becomes a part of our ontology in that setting. However, our classical field equations for example only carry global gauge invariance. Local gauge invariance does not carry through nor is it necessary to tell us about the physical data about our classical equations. 

It is important to reconcile the fact that there are an infinite number of ways we could utilize our gauge freedom in setting up our equations. This brings us to Quine's idea of the proxy function somewhat loosely, in which the various choices of gauge can yield the same correct physical result. There is no "true gauge" specified by even more fundamental principles. In other words, local gauge symmetry in particular can be treated as a proxy for all the various gauge constraints we impose on our equations in the theory where local gauge invariance maps onto physical entities. Therefore, it is not strictly ontological, as global gauge symmetry may be taken to be, rather it is which masks the fundamental principles in lieu of redundancies, but also plays an ontological role in a particular regime. \footnote{As arbiters of Quantum Field Theory, we can make the claim that photons exist independent of whether or not they arise as a result of gauge symmetry. That being said, the mathematical representations of the photon all exhibit this symmetry and thus we  take the abstract entities that map onto physical data to be one and the same.}

We can think of this shifting scale ontology as if we were trying to resolve the pixels of a computer screen. Various details will come in and out of focus as one zooms in and out of the screen. Our ontological commitments must be modified accordingly. If we are committed to principles, those principles will shift and resolve themselves in accordance to what the physical phenomena necessitate.

\section{\label{sec:level1}Conclusions}

The gauge principle as an intermediary has been explored. We have set up a variety of systems, in various contexts and exhibited the importance of the gauge principle in each instance. We have also shown an instance where our additions to the theory, in the case of an auxiliary field, result in no new physical insight whatsoever, exemplifying the difference between the gauge principle and simply adding extra degrees of freedom to our system with the hopes of new physical information. Moreover, in the case of symmetry breaking, our basic example shows that the Higgs mechanism is also a result  of following our nose after realizing the importance of local gauge symmetry. 

A consideration of this principle poses great foundational problems that have yet to be resolved and warrant further exploration. It is an open question if we can extend the idea of intermediaries and shifting scale ontology to a larger system; therefore, it would be worthwhile to explore these ideas as they relate to the broader architecture of the standard model. An application to accidental symmetries and group symmetries in QFT immediately comes to mind as well as  an application to the renormalization group where our equations are fully derived from various scalings in energy.  This is also particularly relevant given the current landscape of theoretical physics in which QFTs are seen as effective field theories only relevant up to a certain energy scale. This has resulted in a long search for the theory that is more fundamental and which will resolve the decades old problem, still withstanding, of quantizing the gravitational force.  Moreover, such an approach will surely have consequences for a broader metaphysical set up which can extend itself into larger epistemological considerations. 

One can conceive of a philosophical system that is sparred by the conception of intermediaries which present themselves as fundamental as various initial conditions are presented.  
\begin{acknowledgments}
We wish to thank Aden Evens and Erkki Wilho Mackey for useful conversations and for reading the initial draft of this work. We also wish to thank Carlo Rovelli for useful clarifications via e-mail correspondence as well as Laura Reutsche for helpful resources as this work was being completed. 
\end{acknowledgments}

\appendix

\section{Spinor Helicity Conventions}

We introduce
		 \begin{equation}
		 	\sigma^\mu = (1,  \sigma^i ), \quad \quad \quad \bar \sigma^\mu =  (1, -\sigma^i),
		 \end{equation}
		 where $\sigma^i$ are the standard Pauli matrices and
		 \begin{equation}
		 	\gamma^\mu =  \left(\begin{matrix} 0 & (\sigma^\mu)_{\dot a  b} \\ (\bar \sigma^\mu)^{ a \dot b} &0\end {matrix} \right) .
		 \end{equation}
		 Here, $\gamma^\mu$ are the usual gamma matrices obeying the Clifford algebra
		 \begin{equation}
		 	\left \{ \gamma^\mu, \gamma^\nu\right\}= -2 \eta^{\mu\nu}.
		 \end{equation}
		Defining
		  \begin{equation}
		 	\begin{split}
		 	p_{\dot a  b}& \equiv \frac{1}{\sqrt{2}}p_\mu(\sigma^\mu)_{\dot a  b} =\frac{1}{\sqrt{2}} \left (\begin{matrix} -p^0 + p^3 & p^1 -i p^2\\ p^1 + i p^2 & -p^0 -p^3
					\end{matrix} \right)
					 , \\ 
		 	 p^{ a \dot b}& \equiv  \frac{1}{\sqrt{2}}p_\mu  (\bar \sigma^\mu)^{ a \dot b}  = - \frac{1}{\sqrt{2}} \left (\begin{matrix} p^0 + p^3 & p^1 -i p^2\\ p^1 + i p^2 & p^0 -p^3 
			\end{matrix} \right)
			 ,
			\end{split}
		 \end{equation}
		 we obtain expressions for null momenta in terms of two-component spinor helicity variables: 
		  \begin{equation}
		    \begin{split}
		 	p_{\dot a b} &= - | p]_{\dot a } \langle p |_b = -\tilde{\lambda}_{\dot a} \lambda_{b},\\ p^{a \dot b} &= - |p \rangle^{a} [p|^{\dot b}=-\lambda^{a}\tilde{\lambda}^{\dot b},
		 	\end{split}
		 \end{equation}

		  Indices are raised and lowered with the Levi-Civita symbol: 
		 \begin{equation}
		 	[p|^{\dot a} = \epsilon^{\dot a \dot b} |p]_{\dot b}, \quad \quad \quad |p\rangle^a = \epsilon^{ab} \langle p |_b
		 \end{equation}
		 where
		 \begin{equation}
		 	\epsilon^{ab} = \epsilon^{\dot a \dot b} = \left(\begin{matrix} 0&1 \\ -1&0\end{matrix}\right).
		 \end{equation} 
		  
		 Finally, in these conventions
		 \begin{equation}
		     p_i \cdot p_j = \langle ij \rangle[ ij ] ,
		 \end{equation}
		 which can be readily verified using the identity
		 \begin{equation}
		     \sigma^\mu_{\dot a a}~\bar \sigma^\nu{}^{ a \dot a} = -2 \eta^{\mu \nu}.
		 \end{equation}

\end{document}